\documentclass{optica-article}

\journal{opticajournal} 

\articletype{Research Article}
\usepackage{amsmath}
\usepackage{textcomp}
\usepackage{amsmath}
\usepackage{siunitx}
\usepackage{float}
\usepackage{graphicx}
\usepackage{textcomp} 
\usepackage{mathptmx}  

\begin{document}

\title{Optical design and characterization of a multi-depth vision simulator}

\author{Parviz Zolfaghari,\authormark{1,*} Ehsan Varasteh,\authormark{1} Koray Kavaklı,\authormark{1} Arda Gulersoy,\authormark{1} Afsun Şahin,\authormark{2,3} and Hakan Urey\authormark{1,2,*}}

\address{\authormark{1}Koç University, Department of Electrical and Electronics Engineering, Istanbul, 34450, Turkey\\
\authormark{2}Koç University Translational Medicine Research Center (KUTTAM), Istanbul, 34450, Turkey\\
\authormark{3}Department of Ophthalmology, Medical School Koç University, Istanbul, 34450 Turkey}

\email{\authormark{*}pzolfaghari@ku.edu.tr; hurey@ku.edu.tr} 


\begin{abstract*}
We present a vision simulator device (Katsim), a compact near-eye optical display designed for assessing postoperative corrected vision, preoperative intraocular lens (IOL) assessment, and objective IOL characterization. The system forms a virtual image using an amplitude-modulated LCoS spatial light modulator (AM-SLM), RGB LED illumination, and a high-speed varifocal lens. In the proposed architecture, the LED illumination and varifocal lens diopter changes are triggered in synchrony with the SLM RGB subframes, rendering three depth planes perceptually simultaneously via high-frequency time-multiplexing. Operating at 60 frames per second (fps), the system achieves an effective 180 Hz depth-coded cycle, enabling sharp, multi-depth rendering within a dynamically adjustable depth range from 0.2 m to optical infinity. The system’s eyebox is configurable from 1 to 5 mm, while maintaining a fixed spatial location and preserving angular magnification regardless of changes in focus or eyebox size. The designed system features a 9.15° field of view. An integrated infrared pupil-tracking module detects non-cataractous regions of the cataractous crystalline lens, and the projected imagery is mechanically steered through those clear zones in real time. The proposed vision simulator supports both subjective simulation of post-surgical vision for patient-specific counseling and objective optical evaluation of IOLs, including resolution and contrast fidelity (e.g., modulation transfer function, contrast transfer function, and defocus curves). By decoupling depth modulation from eyebox position and size, the system offers a modular, portable platform that supports enhanced preoperative planning, personalized IOL selection, objective IOL characterization, and use as a novel research vision tool.

\end{abstract*}

\section{Introduction}
Cataracts are the world’s leading cause of reversible blindness and can progress to complete vision loss if left untreated \cite{bourne2013causes}. The condition arises when the crystalline lens becomes cloudy, scattering and distorting light, which impairs vision, daily functioning, and quality of life, especially in the aging population. Thus, effective surgery, improved diagnostics, and personalized outcome prediction tools are urgently needed \cite{marcos2021simulating}.

Currently, the only definitive treatment for cataracts is surgical extraction of the cloudy lens and implantation of an intraocular lens (IOL). While generally successful, the procedure eliminates the eye’s natural accommodation, placing the optical burden on the IOL. A wide variety of IOLs, including monofocal, bifocal, trifocal, EDOF, and toric types, are available \cite{soomro2024head}. Although multifocal and trifocal lenses aim to provide spectacle-free vision at multiple distances, they may induce halos, glare, or reduced contrast. Preoperative evaluation remains challenging, as selection still depends on ocular biometry, surgeon experience, and basic optical tests, which fail to predict real-world and patient-specific visual performance. Tools such as the pinhole test and Potential Acuity Meter (PAM) are also unreliable in dense cataracts and cannot anticipate postoperative phenomena like glare or halo perception. Postoperative assessments typically measure visual acuity, contrast sensitivity, and through-focus performance \cite{marx2024ability}, but these objective metrics often fail to predict subjective outcomes.

In parallel, objective assessment of IOLs using standardized optical metrics is critical \cite{aydindougan2020applications}. Current gold-standard evaluations use model-eye bench testing with Modulation Transfer Function (MTF), Point Spread Function (PSF), Contrast Transfer Function (CTF), and defocus curves \cite{aydindougan2020applications}, yet these are constrained by simplified ocular models and media. To bridge measurements and perception, visual simulators have been developed \cite{barcala2023visual}, evolving from fixed optics to platforms integrating deformable mirrors, tunable lenses, phase plates, or spatial light modulators (SLMs) for more dynamic and accurate simulations \cite{marcos2022adaptive}.

Within this landscape, there remains a need for content- and depth-adjustable pre- and postoperative visual simulators \cite{aydindougan2020applications}. Complementary approaches include holography-based systems that emulate monofocal and multifocal optics and their effects on contrast and halos \cite{akyazi2024intraocular}; holographic displays for preoperative visual-acuity evaluation \cite{kavakli2021pupil}; geometric-optics simulators for near and distance vision with multifocal IOLs \cite{na2022novel}; and adaptive-optics platforms that project through-focus optical quality and visual acuity at the pupil plane \cite{benedi2021optical}. Display methods that reproduce depth cues, light-field \cite{lanman2013near}, varifocal (tunable-lens) \cite{liu2023recent,zaytouny2023clinical}, and light-sheet displays offer practical routes to depth-continuous preoperative visualization alongside Adaptive Optics (AO) and Computer-Generated Holography (CGH). Concurrently, advances in AR, near-eye displays, and SLMs expand ophthalmic simulation and diagnostics \cite{soomro2024head,paniagua2023economic}, including simulations of cataract and presbyopia outcomes \cite{krosl2020cataract,jones2020seeing}. While VR and CGH are promising for education, most systems remain limited by static content and minimal clinical integration; a notable exception is the pupil-steering CGH near-eye display by Koray et al., which demonstrates depth-adjustable projection \cite{kavakli2021pupil}. Nevertheless, comprehensive assessment of IOL performance that unites optical metrics with subjective visual experience remains challenging.
Visual simulators aid patient education and can accelerate iterative IOL design by enabling pre-fabrication optical evaluation, reducing development time and providing early feedback \cite{marcos2021simulating}. By emphasizing subjective visual experience in clinical validation, preoperative simulators that deliver real-time patient feedback may transform regulatory evaluation frameworks. Commercial systems (OptiSpheric IOL PRO2, PMTF) provide precise MTF and focal-shift measurements but are large, costly lab tools using artificial eyes and narrowband, idealized conditions, limiting ecological validity \cite{wu2025using}. Adaptive-optics simulators can predict retinal images yet are too complex and expensive for broad use, leaving a need for a compact platform that bridges objective IOL assessment and subjective visual simulation under realistic conditions. 

To fill this gap, we present a multi-depth vision simulator: a compact electro-optical platform that unites subjective visual simulation with objective optical performance evaluation, as shown in Fig. 1. Its modular architecture integrates an amplitude-modulated SLM, a tunable varifocal lens, and a real-time bright-pupil tracking module \cite{zolfaghari2025bright}, enabling mechanical eyebox steering to deliver targeted stimuli through clear regions of the crystalline lens or a model eye, which is critical for assessing partial opacities and localized cataracts.

Our device projects depth-adjustable, high-resolution 3D content through the pupil with synchronized RGB illumination, each channel encoding a focal plane. The varifocal lens is synchronized with SLM updates and RGB timing, enabling \textit{in situ} simulation of refractive errors (myopia, hyperopia). A key innovation is the ability to vary focus and eyebox size via magnification and diffuser-spot control, without shifting eyebox location or changing angular magnification; this is achieved through optical optimization. This stability preserves consistent conditions for both subjective assessments and objective lens characterizations.
Clinically, Katsim serves as a patient-facing simulator for preoperative counseling, with personalized visualizations of monofocal, EDOF, trifocal, and toric lenses under realistic conditions. It accommodates patient-specific pupil size/position and refractive profile, providing an immersive, tailored preview of outcomes. In parallel, Katsim is a research-grade evaluation platform, projecting calibrated test patterns through model eyes (commercial or prototype IOLs), enabling quantitative MTF, CTF, defocus-curve, and PSF measurements under realistic conditions (decentration, pupil motion, off-axis aberrations) that are difficult to reproduce on conventional benches.
Katsim’s compact, modular design suits outpatient clinics, pre-surgical consultations, and research laboratories. As one of the first ophthalmic systems with dual-mode operation, subjective simulation and objective lens testing, it unifies patient-specific visualization and optical metrology under consistent conditions, advancing personalized medicine, accelerating IOL development, and supporting simulation-based regulatory evaluation.

\begin{figure}[!t]
\centering
\includegraphics[width=5.2in]{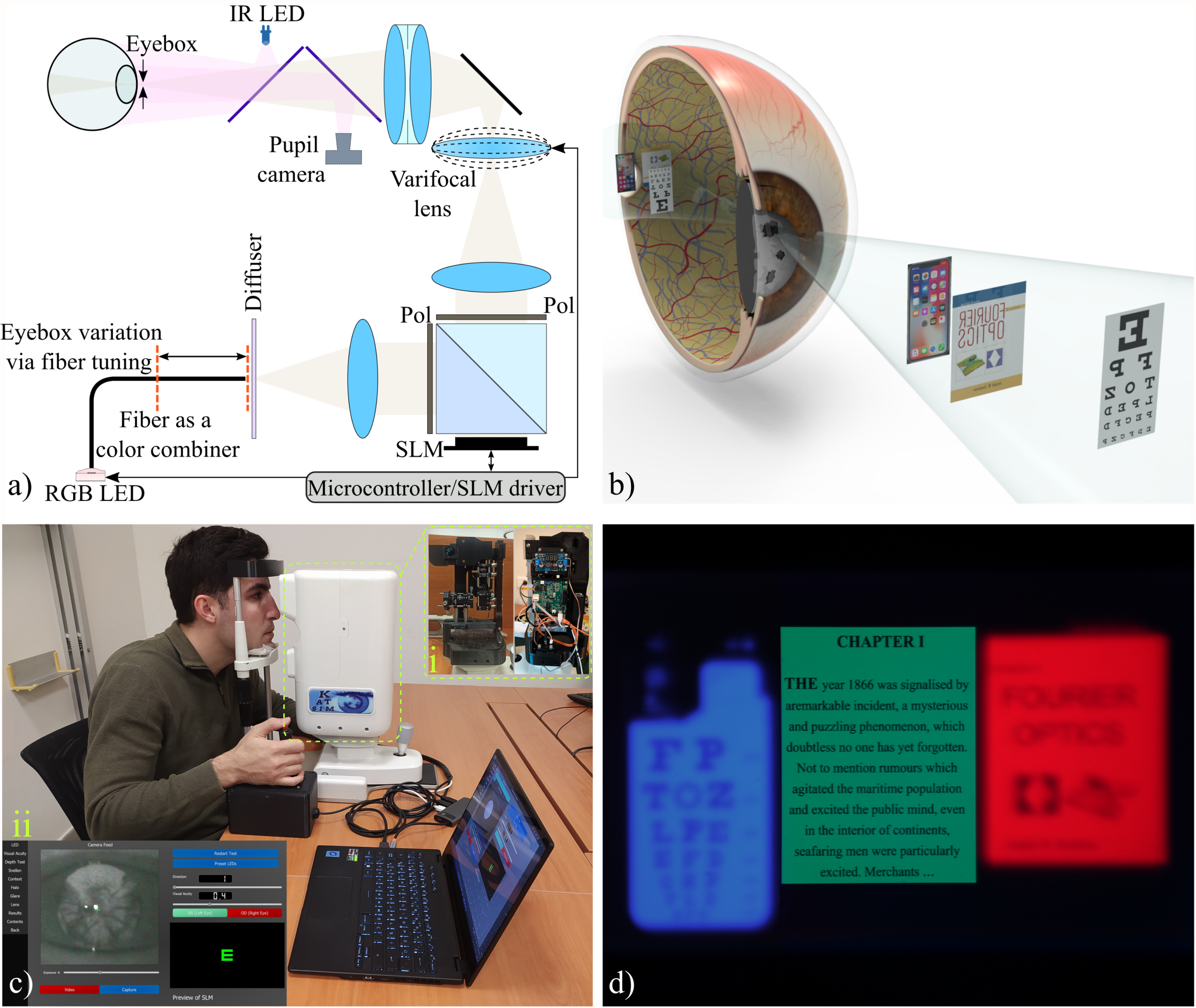}
\caption{Overview of the KATSIM depth-adjustable near-eye display system. (a) Optical schematic of the KATSIM setup, incorporating an SLM and a varifocal lens; Pol: Polarizer. (b) Illustration of multi-depth-resolved content projected onto the retina. (c) A subject undergoing a visual test with KATSIM. Insets: (i) optical image of the proposed electro-optical hardware; (ii) human–machine interface software screen for multifunctional control of KATSIM, showing the patient’s cataractous crystalline lens captured by the pupil tracker. (d) Content image captured by a camera at the eyebox plane, focused at 1 m, illustrating the corresponding retinal image. See Visualization 1 for a demonstration of the three-depth operation and the eyebox-plane camera capture.}
\label{fig:1}
\end{figure}

\section{Method}
\subsection{System Concept and Architecture}
Advances in augmented reality (AR), near-eye display systems, and SLM technologies have enabled the development of novel ophthalmic simulation platforms. These tools are particularly valuable for assessing visual performance in patients with cataracts or other intraocular optical impairments. In this study, we introduce Katsim, a modular, compact, and depth-adjustable near-eye display system designed for cataract simulation and preoperative evaluation of intraocular optical quality, including potential compatibility with IOLs. The system supports both subjective visual simulation for patient counseling and objective optical characterization of IOLs, providing a dual-mode platform for clinical and research use.

As illustrated in Figure 1a, the Katsim system integrates three core modules: a commercial AM-SLM for high-resolution image projection; a focus-tunable varifocal lens (Optotune EL-3-10) for real-time depth modulation; and a bright-pupil infrared eye-tracking unit for dynamic monitoring of pupil geometry and crystalline lens opacity. The AM-SLM is a microdisplay with a pixel pitch of 4.5 µm and operates at 60 frames per second. It electronically modulates the light intensity on a pixel-by-pixel basis to generate customizable visual content using an RGB light source, including optotypes, natural scenes, or structured wavefront patterns. This flexibility allows for simulation of a wide range of visual stimuli under realistic ophthalmic conditions. The varifocal lens follows in the optical path and provides electronically tunable focal length control by modulating its internal curvature. This enables smooth, continuous shifts in the virtual image plane across a range of distances—from near (25 cm) to far (4 m or infinity). To enable high-speed performance, a custom low-pass filter is implemented in the microcontroller, reducing the settling time of the varifocal lens to under 2 milliseconds. This ensures rapid transitions during dynamic depth rendering, which is critical for simulating natural accommodation and for correcting refractive errors such as myopia or hyperopia.

Unlike conventional near-eye displays with fixed optical paths and limited eyebox sizes, Katsim features a mechanically adjustable eyebox (exit pupil) size. As illustrated in Figure 1a, the diffuser and the eyebox plane are optically conjugated. By mechanically varying the distance between the fiber-optic light source and the diffuser, the size of the illumination spot on the diffuser changes, which in turn alters the eyebox size. Through careful optical optimization, this change in magnification and focus does not shift the eyebox location or alter the angular magnification, ensuring consistent image geometry on the retina. This mechanically tunable configuration allows the system to dynamically adapt the eyebox to match the patient’s pupil size, simulate realistic visual effects under varying optical conditions, and facilitate IOL evaluation under stable and repeatable conditions.

To identify clear optical zones, the system employs a real-time infrared (IR) bright-pupil eye-tracking module, consisting of a co-axially aligned IR-LED and camera. In this configuration, reflections from the retina appear bright through transparent areas of the crystalline lens, whereas cataractous or opaque regions appear dark. By capturing live video of the pupil and crystalline lens, the system dynamically maps the spatial distribution of lens opacity. As illustrated in Figure 1b, the eyebox is then mechanically steered to direct the projected light through the clearest portion of the lens, thereby enhancing retinal image quality. This targeted steering goes beyond conventional eyebox expansion by enabling spatially selective image delivery—an essential feature for realistic simulation in optically degraded eyes.

For clinical usability and measurement repeatability, the entire optical-electronic assembly is mounted on a standard ophthalmic examination stand with an integrated headrest, as shown in Figure 1c. This setup stabilizes the patient’s head during testing and allows clinicians to monitor the pupil and lens condition in real time via an external display.

The Katsim system is capable of projecting depth-encoded, high-resolution RGB content at multiple virtual distances, specifically near (25 cm), intermediate (60 cm), and far (4 m), as shown in Figure 1d. When the imaging camera is focused at the intermediate virtual distance (60 cm), objects rendered at near and far depths appear appropriately blurred, thereby replicating realistic depth-of-focus effects and providing natural accommodation cues. In addition to clinical simulation, Katsim enables objective evaluation of IOL performance, including defocus curves and CTF measurements, by projecting structured stimuli through model eyes. The proposed system is particularly suited for assessing visual performance in eyes with partial crystalline lens opacities and holds potential for simulating IOL placement strategies and outcomes prior to surgery.

\begin{figure}[H]
\centering
\includegraphics[width=5.2in]{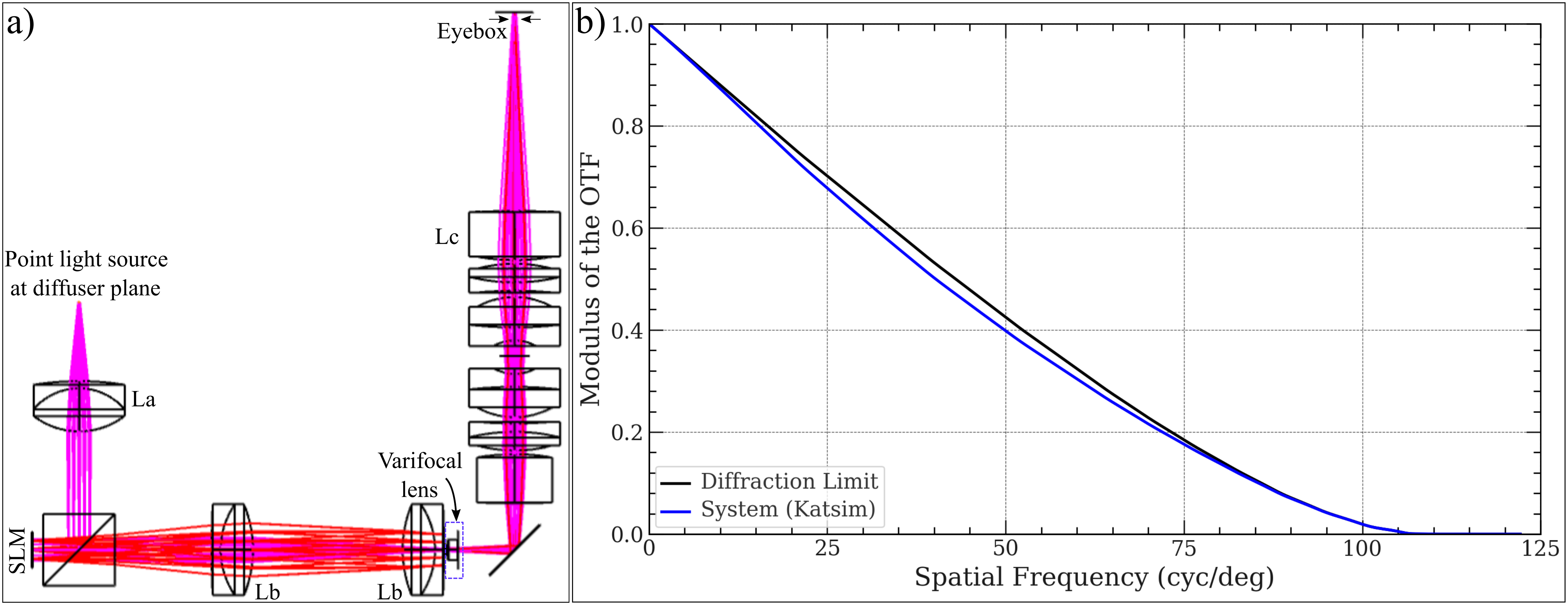}
\caption{(a) Zemax optical layout of the system, showing the SLM, varifocal lens, and eyebox-projection; incorporating lenses: La = AC254-030 and Lb = AC254-050 (Thorlabs), Lc = 6-element Edmund Optics lens (part number 45760).
(b) Simulated CTF of proposed system (blue) compared with diffraction-limited performance (black). The simulation was performed for a 3 mm eyebox diameter.}
\label{fig:2}
\end{figure}

\subsection{Optical Design}
The optical design and ray-tracing simulation of the Katsim system were conducted using Zemax OpticStudio, as illustrated in Figure 2a. The optical configuration includes a collimating lens (AC254-030, focal length: 30 mm), followed by a 4f relay system comprising two 50 mm focal length lenses (AC254-050) with a varifocal lens (Optotune EL-3-10). A 50 mm focal length lens is positioned directly in front of the varifocal lens, serving a dual purpose: it increases the effective dioptric power range of the varifocal lens and shortens the overall optical path, thereby optimizing the system’s compactness and form factor. For the eyepiece, the ray-tracing simulation employs a 6-element Edmund Optics lens (part no. 45760) to model retinal image formation. However, in the experimental prototype, a Nikon 50 mm f/1.8D lens is used due to practical availability in the laboratory. As shown in Figure 2a, the system aligns several key conjugate planes to maintain accurate light propagation and depth simulation: the diffuser plane, varifocal lens plane, and eyebox plane are optically conjugated; likewise, the SLM plane, eyepiece lens aperture, and retinal plane form another set of conjugate surfaces.

A notable feature of the design is that adjusting the varifocal lens diopter alters the virtual image depth without changing the location or size of the eyebox or affecting angular magnification. This invariance was achieved through careful optical optimization of the conjugate planes and magnification system. It ensures that the projected image enters the eye consistently through the correct pupil location, regardless of the rendered depth. The calculated field of view (FOV) of the system is 9.15 degrees, suitable for ophthalmic simulation tasks. Optical Transfer Function (OTF) analysis for a 3 mm pupil size and a point light source at the diffuser plane is presented in Figure 2b. The simulated OTF remains above 0.5 at spatial frequencies exceeding 30 cycles/degree, surpassing the typical resolution limit of human vision and confirming that the system can support high-resolution visual simulations.

To render realistic 3D content, the system dynamically adjusts the focal length of the varifocal lens, shifting the virtual image plane across a range of depths. The conjugate alignment of the optical components not only facilitates depth modulation but also enables independent mechanical tuning of the eyebox size without affecting the eye relief. Importantly, the system maintains a fixed eyebox location and angular image properties during both magnification and focus changes, supporting consistent projection geometry during depth simulation. This is particularly beneficial for accommodating different pupil sizes, minimizing light scatter, and maintaining stable image quality—especially important when evaluating vision in the presence of IOLs or other ocular conditions. Optimizing the eyebox size is crucial for user comfort, robust alignment, and ensuring consistent image delivery through the clearest regions of the eye. These characteristics support the system’s dual purpose: enabling subjective visual acuity simulation in cataract patients, and performing objective IOL evaluation using defocus curve and CTF measurements.

\begin{figure}[H]
\centering
\includegraphics[width=4.8in]{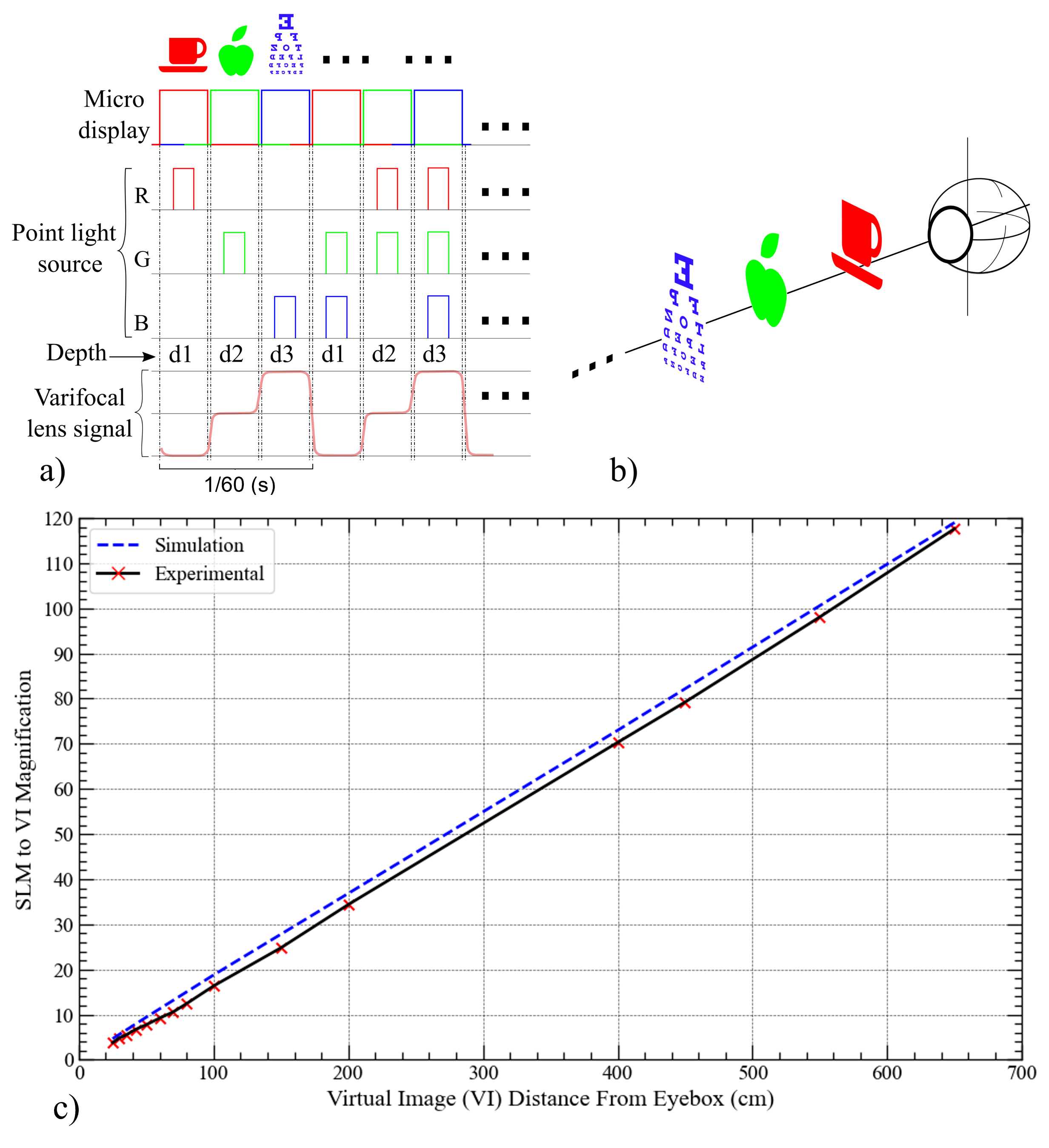}
\caption{(a) Temporal synchronization of RGB subframes with varifocal lens modulation enables full-color 3D image generation by assigning distinct depths (d1, d2, and d3) to each color channel within a single frame.
(b) Conceptual illustration of the three depths corresponding to RGB subframes.
(c) System magnification versus virtual image (VI) distance, showing strong agreement between experimental measurements and simulation.}
\label{fig:3}
\end{figure}

\subsection{System's Electro-Optical Synchronization}
The Katsim system generates full-color, depth-resolved 3D visual stimuli by tightly synchronizing its AM-SLM, electrically tunable varifocal lens, and custom-controlled RGB LED light sources, as shown in Figure 3a. Each video frame, operating at 60 frames per second (fps), is composed of three color channels, red, green, and blue, each encoding a distinct visual scene corresponding to a specific virtual depth plane (e.g., far, intermediate, near). These color channels are temporally multiplexed and sequentially displayed at an effective rate of 180 Hz.

During each subframe (R, G, or B), the varifocal lens dynamically adjusts its dioptric power to shift the focal plane of the optical system to the corresponding virtual depth. Simultaneously, only the relevant corresponding RGB LED (red, green, or blue) is activated to illuminate the content being written to the SLM for that channel. This temporally multiplexed approach allows the system to project three spatially distinct image layers, each corresponding to one channel of the same frame, within a single frame cycle, as illustrated in Figure 3b. The brightness and color accuracy of each plane are further refined through modulation of the LED drive current. A central microcontroller governs the synchronization of all components. Specifically, the AM-SLM is driven by a dedicated SLM driver board that generates frame refresh signals and channel-specific control pulses for the red, green, and blue subframes. These signals serve as the synchronization backbone for the entire system. The microcontroller receives these channel pulses and coordinates the corresponding focal shifts of the varifocal lens and the switching and intensity modulation of the customized RGB LEDs. A low-pass filter is integrated on the main control board to minimize the settling time of the varifocal lens to less than 2 milliseconds, ensuring that focal transitions are completed before illumination occurs. Each LED is activated only after the varifocal lens stabilizes at the correct focal power and is turned off prior to advancing to the next subframe, thereby preventing depth crosstalk or motion blur, as illustrated in Figure 3a.

This architecture offers flexible and precise control over both the virtual depth and visibility of each projected image layer. Depth planes can be independently toggled by selectively enabling or disabling their corresponding LEDs, while their virtual distances are modified by adjusting the varifocal lens diopter. Furthermore, LED intensity modulation allows for accurate color reproduction and visual emphasis at each depth. As a result, the system is capable of presenting patient-specific 3D scenes (such as optotypes at 4 m, 60 cm, and 25 cm) within a single frame cycle, and aligning each projection through optically clear zones of the eye, even in cases of cataract-induced opacity. The ability to shift focal depth without affecting eyebox location or angular magnification is maintained throughout this process, a feature achieved through careful optical design and ray-tracing optimization.

To validate the optical behavior of the system, specifically the impact of varifocal lens diopter adjustments on virtual image plane positioning and magnification, the magnification from the SLM to the virtual image plane was analyzed as a function of virtual image distance, as shown in Figure 3c. This evaluation was conducted both through ray-tracing simulations in Zemax OpticStudio and through experimental measurements using custom calibration patterns. The results confirmed a constant angular magnification, indicating that the system preserves expected optical scaling behavior across varying depths. This optical consistency is critical for both preoperative subjective simulation of visual acuity in cataract patients and objective evaluation of IOLs, including defocus curves and CTF measurements. These results support the system’s suitability for simulating multifocal and extended depth-of-focus IOLs and for use in preoperative visual performance assessments.

\begin{figure}[H]
\centering
\includegraphics[width=5in]{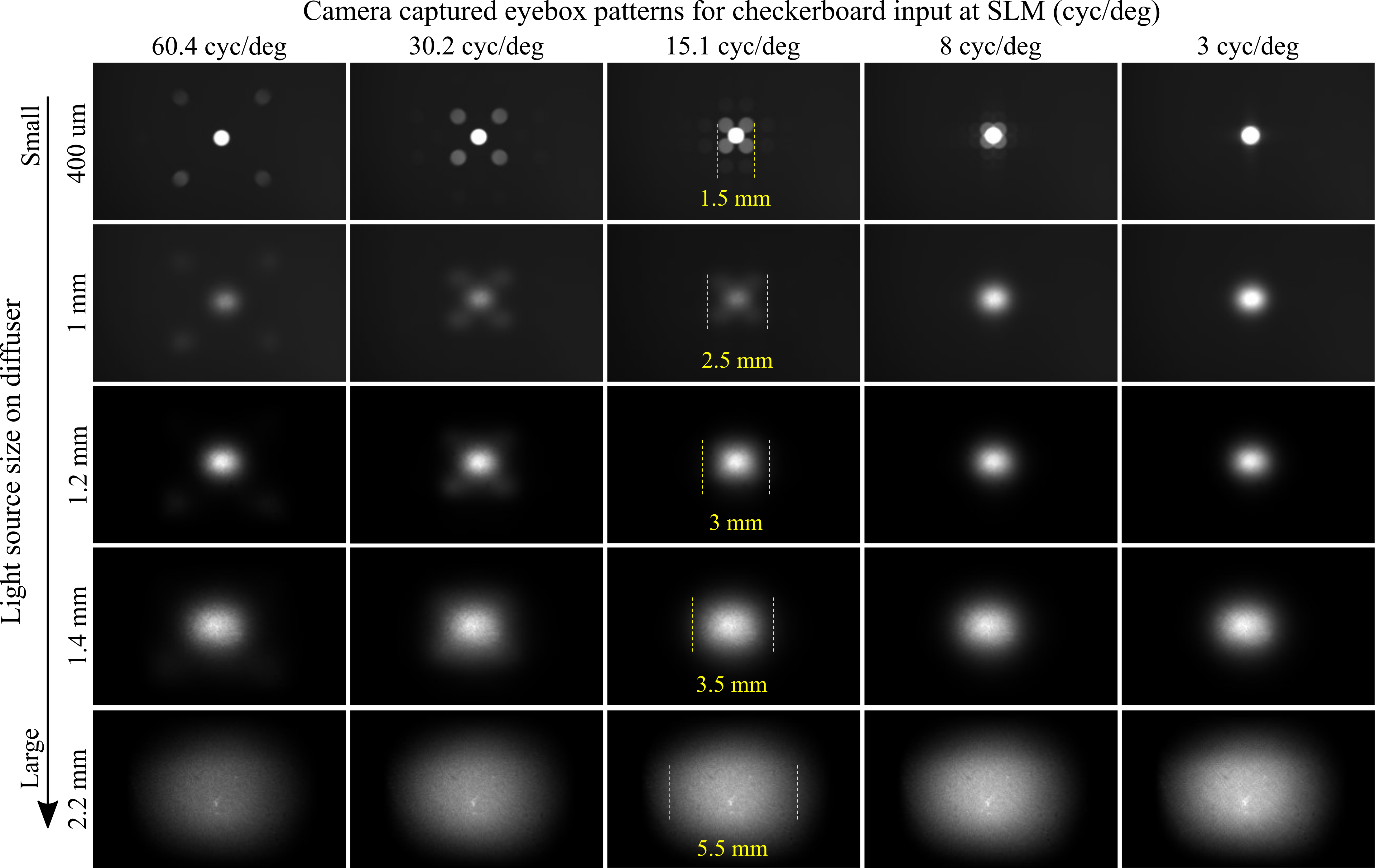}
\caption{Camera-captured eyebox patterns for different light-source diameters (0.4–2.2 mm) and checkerboard inputs on the SLM illustrate how source size influences modulation transfer across spatial frequencies. The camera sensor (without additional imaging optics) was placed at the eyebox (exit-pupil) plane. For a 3 mm pupil (aperture), low-frequency patterns transmit efficiently, whereas high-frequency components are partially filtered because their higher-order diffracted beams lie outside the aperture, an effect that is more severe with cataract-level apertures ($\sim$1\,mm).}
\label{fig:4}
\end{figure}

\subsection{Adjustable Eyebox Implementation}

In this subsection, we evaluate the relationship between light source size and exit pupil (eyebox) diameter using the dynamic eyebox scaling capability of the proposed device. The eyebox scaling was evaluated by systematically varying the illumination spot size on the diffuser, which was achieved by adjusting the distance between the optical fiber and the diffuser surface. To observe the eyebox at the Fourier plane, we used a lensless camera positioned at the eyebox plane to capture the projected patterns. As illustrated in Fig.~\ref{fig:4}, the diameter of the eyebox was linearly scaled with the size of the input light source, ranging from approximately $400\,\mu$m to $2.2$~mm. Both experimental data and simulation results yielded a consistent magnification factor of approximately $2.13\times$ between the diffuser and eyebox planes, validating the expected behavior of the system's fixed conjugate-plane design. 

Here, we show how the eyebox diameter increases with the illumination spot size on the diffuser, from about 400 µm to 2.2 mm. Both experimental data and Zemax simulations confirm a fixed magnification factor of 2.13×, consistent with the conjugate-plane design. This stability means the system can adapt to different pupil sizes while keeping alignment consistent, which is critical for patient-specific testing. With this capability, the device can also simulate how implanted IOLs function under mesopic or scotopic conditions (at night), when the pupil is larger than during the day. Importantly, dynamic adjustment of the varifocal lens does not affect eyebox size or location, and eyebox scaling through fiber–diffuser distance changes does not alter system magnification.

However, the size of the eyebox is not only determined by the light source size but also by the spatial frequency content of the projected patterns. From an optical standpoint, the eyebox plane represents the geometrically relayed interaction of spatial illumination and angular modulation. The illumination profile from the diffuser multiplies by the angular content encoded by the SLM, resulting in a mapped superposition at the eyebox plane. Therefore, we recorded checkerboard patterns at the eyebox plane to analyze how spatial frequency content propagates through the system under varying illumination conditions.

We have assessed different spatial frequencies (in cycles per degree) of the checkerboard stimuli displayed on the AM-SLM. As shown in Fig.~\ref{fig:4}, when the light source spot was small, the resulting eyebox was also narrow. High spatial frequency stimuli (e.g., 30.2 or 60.4 cycles/degree) produced high-frequency diffraction orders that were angularly distant from the central (low-frequency) components. Assuming a typical pupil diameter of $5$~mm, the eye effectively acts as an aperture that blocks higher-order diffraction components, thereby reducing the resolution of fine image details. This effect becomes particularly important in patients with narrow, non-cataractous regions of the crystalline lens or inherently small pupils. In such cases, spatially detailed features become difficult to perceive, even when presented at full system resolution.

Conversely, increasing the light source size produces a larger eyebox, allowing more angular orders to enter the pupil. However, if the non-cataractous region is still small, this broader angular spread results in a loss of resolution due to partial clipping of high-frequency components. These observations demonstrate that virtual image depth modulation, controlled by the varifocal lens, does not influence eyebox size or location. This decoupling is consistent with the system’s fixed conjugate-plane architecture, wherein the diffuser plane, varifocal lens, and eyebox plane are optically conjugated. This behavior is a direct result of optical optimization during the system's design phase, which preserves both angular magnification and spatial alignment across depth changes. The ability to independently control depth and eyebox geometry is a key feature that enables flexible, patient-specific visual testing.

Clinically, dynamic eyebox control allows the simulation of restricted viewing conditions, such as narrow pupils or small clear zones in cataract patients. This capability supports preoperative assessment of IOL performance by presenting high-resolution stimuli through the most optically clear region of the eye while minimizing stray light exposure elsewhere in the pupil and reducing the impact of optical aberrations. In addition to subjective visual simulation, this configuration also supports objective evaluation of IOL performance, including CTF and defocus curve measurements, making it suitable for both clinical and research applications.

\begin{figure}[H]
\centering
\includegraphics[width=5in]{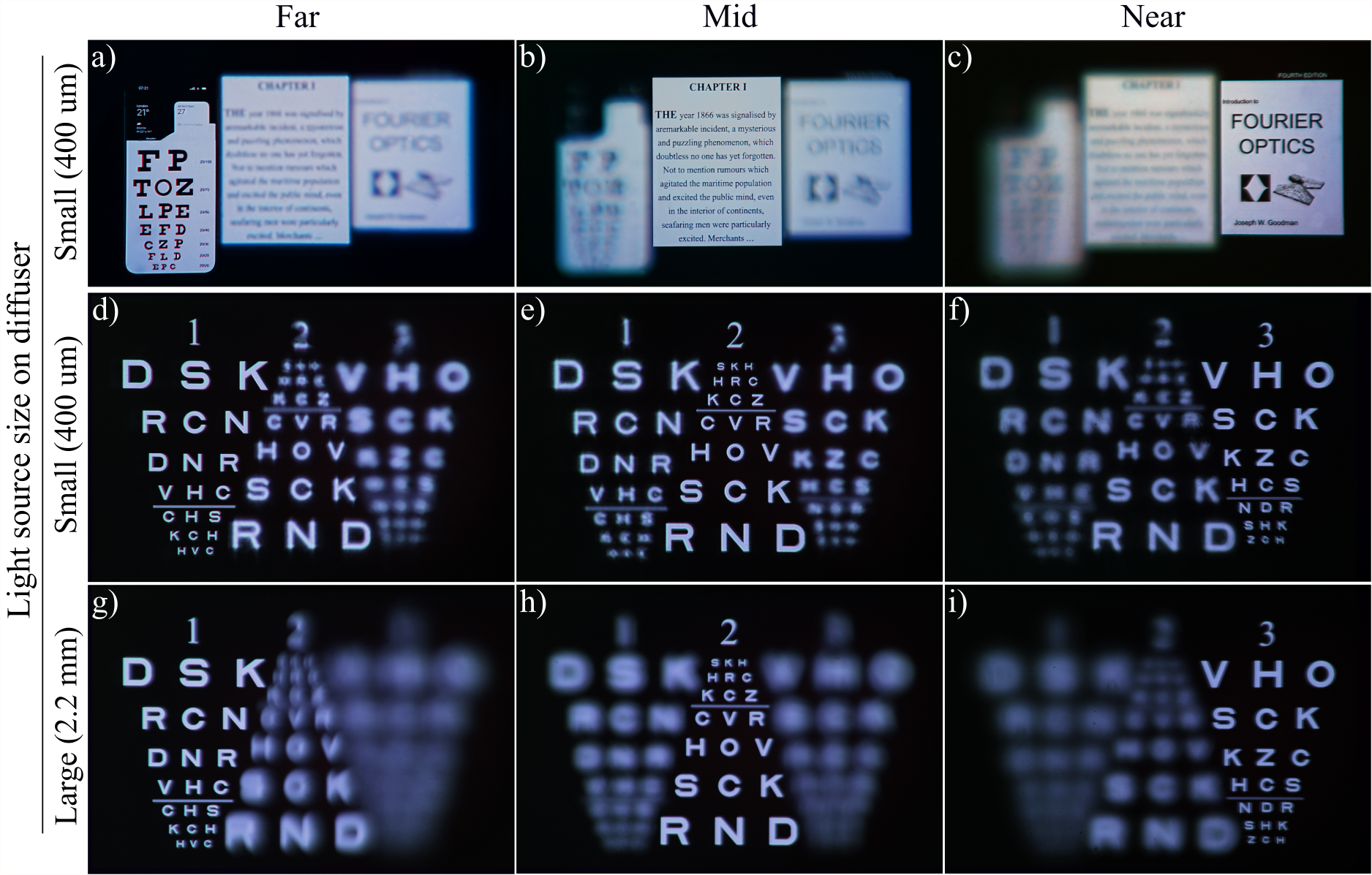}
\caption{Comparison of small (400\,\textmu m) and large (2.2\,mm) illumination-source diameters on the diffuser for black-and-white content at three depths:
far (4\,m; panels a,d,g), intermediate (60\,cm; panels b,e,h), and near (25\,cm; panels c,f,i).
Panels a--f use the small source; panels g--i use the large source.
In all images, the camera was positioned at the eyebox (exit-pupil) plane, and the camera lens was focused sequentially at far, intermediate, and near distances.
A larger source expands the eyebox and enhances perceived depth stratification, whereas a smaller source increases depth of field, resulting in sharper focus across multiple planes. See Visualization 2 for the effect of varying the illumination diameter on perceived depth stratification; see Visualization 3 for independent RGB-intensity control of each depth plane and varifocal-lens dioptric tuning enabling independent depth-plane adjustment from 25 cm to optical infinity.}  
\label{fig:5}
\end{figure}

To evaluate the interplay between light-source size, eyebox size, and virtual 3D imagery, we presented random content and logMAR acuity charts under two illumination conditions on the diffuser: a small spot ($400\,\mu$m) and a large spot ($2.2$~mm) (Fig.~\ref{fig:5}). In all experiments, imagery was captured with a synchronized camera; for virtual-plane capture, the camera focus was set to the target far ($4$~m), intermediate ($60$~cm), or near ($25$~cm) depth, and acquisition was gated to the corresponding RGB subframe. With the small illumination, Fig.~\ref{fig:5}a--c show random content at far, intermediate, and near, and Fig.~\ref{fig:5}d--f show the corresponding logMAR charts. With the larger illumination, Fig.~\ref{fig:5}g--i present logMAR charts at the same three depths. Increasing the illumination footprint, and thus the eyebox diameter, made depth stratification more apparent: when focused at one plane (e.g., $60$~cm), the other planes exhibited appropriate defocus while the in-focus plane retained high spatial fidelity. Conversely, the small-spot condition produced a smaller eyebox and higher on-axis contrast, but restricted peripheral angular content, which can limit access to high-spatial-frequency detail, particularly for small pupils or eyes with localized clear windows. Overall, these results confirm that eyebox scaling is linear with illumination size and remains decoupled from varifocal depth modulation, in agreement with the conjugate-plane design.

\section{Results}

\subsection{Katsim optical characterization}
We employed a camera-based imaging system to measure an optical quality metric and thereby quantify the optical performance of proposed system. As shown in Fig. 6b, a high-resolution scientific camera was positioned directly in front of the viewing optics, and the proposed device was mounted on a fixed optical bench. This setup enables precise characterization of spatial resolution and contrast transfer behavior.

Rather than using a standard USAF 1951 resolution target, we designed a custom digital test target composed of bar-chart regions spanning a range of spatial frequencies (see Fig. 6a). The test patterns were displayed by KATSIM and captured by the camera. Each bar region corresponds to a distinct spatial frequency band, allowing measurement of the CTF over a broad frequency range under realistic imaging conditions.

\begin{figure}[H]
\centering
\includegraphics[width=5in]{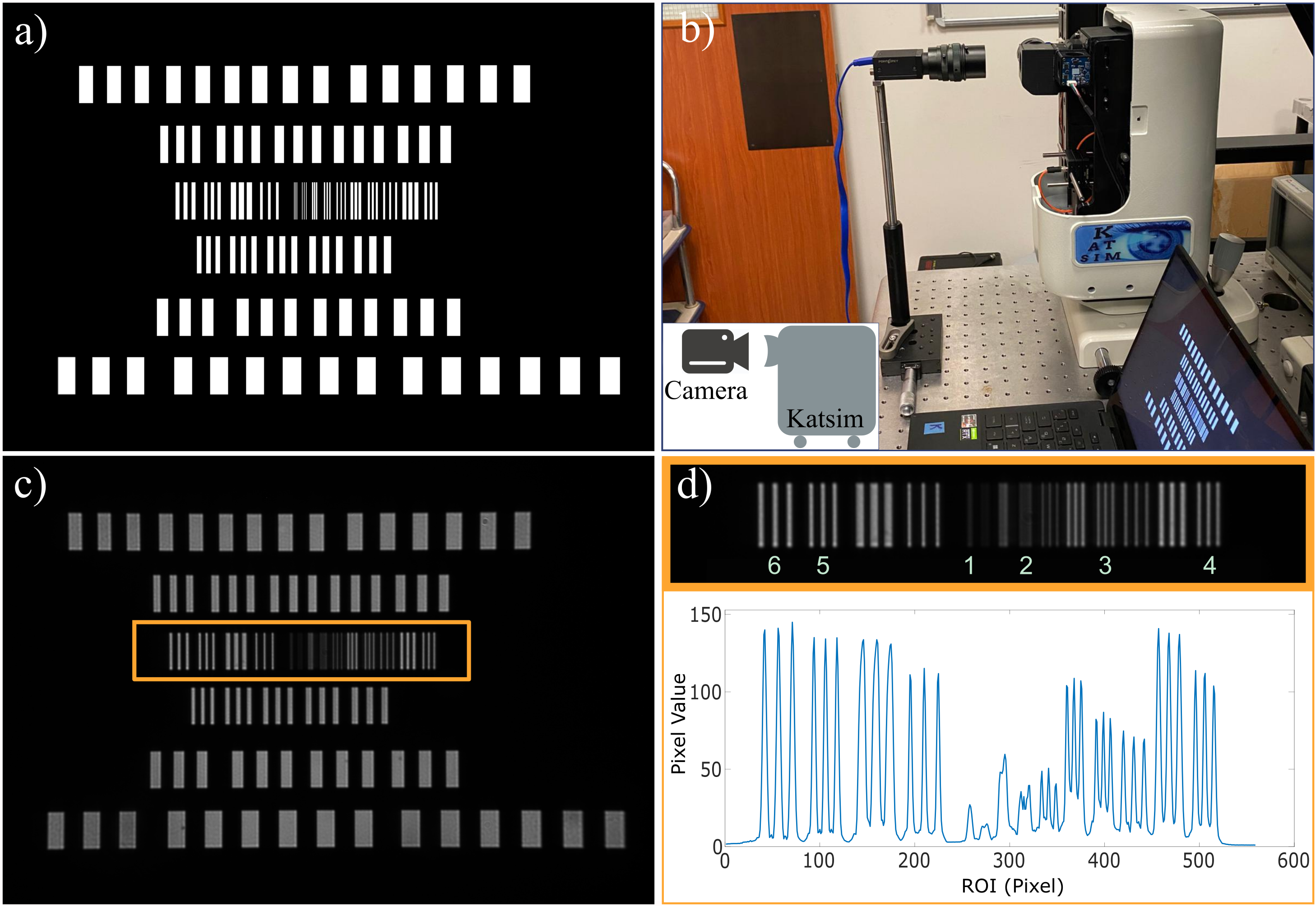}
\caption{Katsim System optical performance for different characterization. (a) Customized spatial-frequency patterns displayed on the SLM.
(b) The camera was positioned at the exit-pupil (eyebox) plane for image capture.
(c) Example capture with the camera lens focused to a representative virtual depth (here, 42\,cm).
(d) High-spatial-frequency patterns with corresponding cross-sections ranging from 2~pixels per cycle to 3~pixels per cycle, with ROI pixel-intensity values. Numbers indicate pixel-pair counts: ``1'' = 1 pair (2-pixel stripe: 1~on, 1~off), ``2'' = 2 pairs (4-pixel stripe: 2~on, 2~off), and so on.}
\label{fig:6}
\end{figure}

Fig. 6a shows the primary stimulus used to evaluate the optical performance of the proposed system, comprising digitally generated bar-chart regions, each corresponding to a distinct spatial frequency. The highest frequency band uses alternating 1-pixel white / 1-pixel black bars (1:1 duty cycle). Subsequent bands reduce the spatial frequency by increasing stripe width in integer steps, up to bars of 60 pixels in width. To ensure uniform sampling by the camera and mitigate size-dependent bias, all frequency bands were rendered at the same bar-chart height.

To further probe the fidelity of contrast reproduction under nonuniform periodic patterns, we introduced an additional region at the high–spatial-frequency end (from 1:1 to 5:5) containing asymmetric bar charts. In this region, the duty cycles deviate from 1:1; we used patterns of 1:2 (1 px ON, 2 px OFF), 2:1, 2:3, 3:2, 3:4, 4:3, 4:5, and 5:4 to simulate nonuniform periodic features. This asymmetric segment is highlighted by a yellow rectangle in Fig. 6c, indicating its placement within the highest-frequency region. The purpose is to assess how well KATSIM reproduces contrast across different duty cycles - a known factor impacting perceived sharpness and contrast, especially in diffractive systems.

Beyond this asymmetric segment, the remaining symmetric bar patterns span a line-pair range from 8 px to 60 px, with spatial frequency decreasing accordingly. This carefully structured test stimulus enables precise quantification of the CTF over a clinically relevant frequency range, allowing reliable benchmarking of KATSIM’s optical fidelity consistent with prior optical simulation platforms.

To ensure measurement consistency, camera parameters (gain, exposure, and white balance) were held constant throughout acquisition. From each recorded image, intensity profiles were extracted along lines perpendicular to the bar patterns. Within each spatial-frequency region, the maximum and minimum pixel intensities—denoted \(I_{\max}\) and \(I_{\min}\)—were identified. The Michelson contrast for each frequency \(f\) was then computed via

\[
C(f) = \frac{I_{\max}(f) - I_{\min}(f)}{I_{\max}(f) + I_{\min}(f)},
\]

and the resulting contrast values \(C(f)\) were plotted versus spatial frequency \(f\) to form the system’s CTF curve. This procedure delivers a quantitative measure of KATSIM’s spatial resolution and optical fidelity, facilitating comparison with benchmark data from physical IOLs. In Fig. 6c, a rectangular ROI is highlighted; the corresponding region is shown in Fig. 6d, where its pixel-intensity cross-section is plotted below.

The spatial frequencies of each bar pattern were expressed in cycles per degree (cyc/deg). Rather than re-deriving magnification for every test, we refer to the system magnification vs. virtual image (VI) distance curve shown in Fig. 3c. At the VI plane used in our measurements, the effective magnification is known. Using that magnification and the SLM’s pixel pitch (\SI{4.5}{\micro\meter}), we computed the angular subtense of each stripe width and converted stripe widths (in pixels) into angular spatial frequencies. The system’s FOV was computed to be 9.15°, which ensures that the projected image properly enters the eye across the pupil for all rendered depths. 





The CTF results of the device are presented under two conditions: small (400\,\textmu m) and large (2.2 mm) illumination- source sizes. The test pattern was placed at a distance of 50 cm in the KATSIM system, and the contrast for each spatial frequency (i.e. bar width) was computed. As mentioned above, for spatial frequencies corresponding to fractional pixel widths (between 1 and 5 pixels), we generated multiple bar configurations to approximate these non-integer cases. As shown in Fig. 6d, to represent 1.5 pixels, we used both a 1 px ON / 2 px OFF pattern and a 2 px ON / 1 px OFF pattern. For each such frequency, the contrast values from both variants were averaged to derive a representative contrast. For all integer widths above 5 pixels, a single bar configuration was used per frequency.


\begin{figure}[H]
\centering
\includegraphics[width=5.2in]{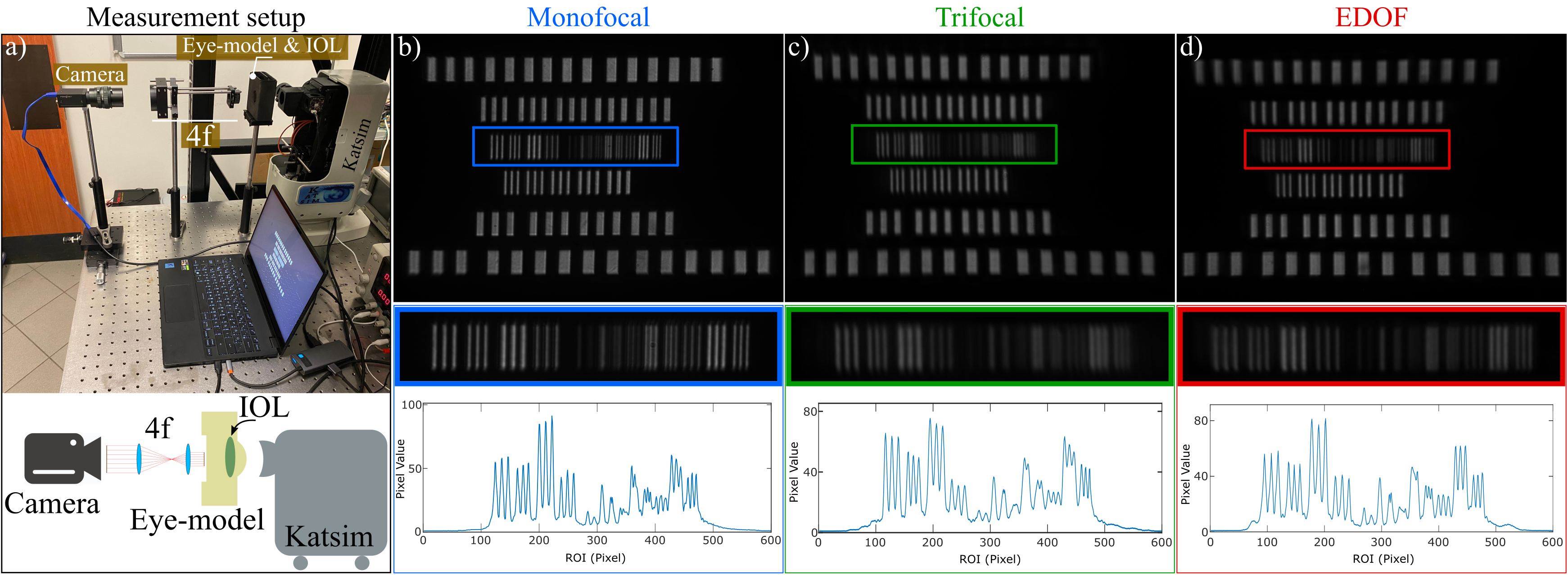}
\caption{IOL characterization using the KATSIM system.
(a) Experimental setup including the KATSIM device, an artificial‑eye model with an implanted IOL, a 4f relay optical system, and a high‑resolution camera.
Spatial‑frequency test patterns captured sequentially with a monofocal IOL (b), a trifocal IOL (c), and an EDOF IOL (d) in the model eye. For the highlighted zones in (b, c, d), the corresponding pixel-intensity cross-sections are plotted below.}
\label{fig:7}
\end{figure}

\subsection{IOL Characterization using KATSIM}

To objectively evaluate the optical performance of IOLs with KATSIM, we used a 3D-printed artificial eye model \cite{akyazi2023artificial} (Fig. 7a). The model comprises a polymethyl methacrylate (PMMA) scleral lens having +4 D optical power (functioning as an artificial cornea), an IOL immersed in liquid, and a glass lamella acting as the retina. The model pupil size was fixed at 6 mm in all experiments; only the Katsim eyebox size was varied to emulate different viewing conditions. The optical layout was designed in Zemax \cite{akyazi2023artificial}, with internal distances of 7 mm from cornea to IOL and 19.7 mm from IOL to retina, yielding a total optical path length of 20.7 mm. 

We tested three IOL types - monofocal (L1), EDOF (L2), and trifocal (L3) - mounted in the artificial eye. For each lens, we first adjusted focus to its primary focal plane and acquired the bar chart image at that plane. We then computed the CTF from those focused images, using the same contrast-versus-frequency analysis methodology as for the KATSIM system itself. The focused captures used for CTF extraction are shown in Figs. 7b–d, and the resulting CTF curves (for small vs. large eyeboxes) are plotted in Figs. 8(a) and 8(c). 

To evaluate through-focus performance of each IOL, we exploited KATSIM’s ability to shift the virtual image depth without physical trial lenses. First, the bar chart stimulus was set such that it produced a focused image on the retina when the virtual object distance was 4 m (0 D defocus), as shown in Figs. 8b and 8d. At that position, we measured the CTF at 30 cycles/degree, a frequency chosen for its sensitivity to defocus degradation.

Next, we simulated defocus by shifting the virtual image plane in ±0.5 D increments above and below 0 D. At each defocus step, we captured an image and extracted the CTF at 30 cyc/deg to build a defocus response curve for each lens. This approach enables efficient, lens-free through-focus testing while maintaining precise control over simulated object depth. The results are presented in Figs. 8b and 8d for small and large eyebox conditions, respectively.

\begin{figure}[H]
\centering
\includegraphics[width=5.3in]{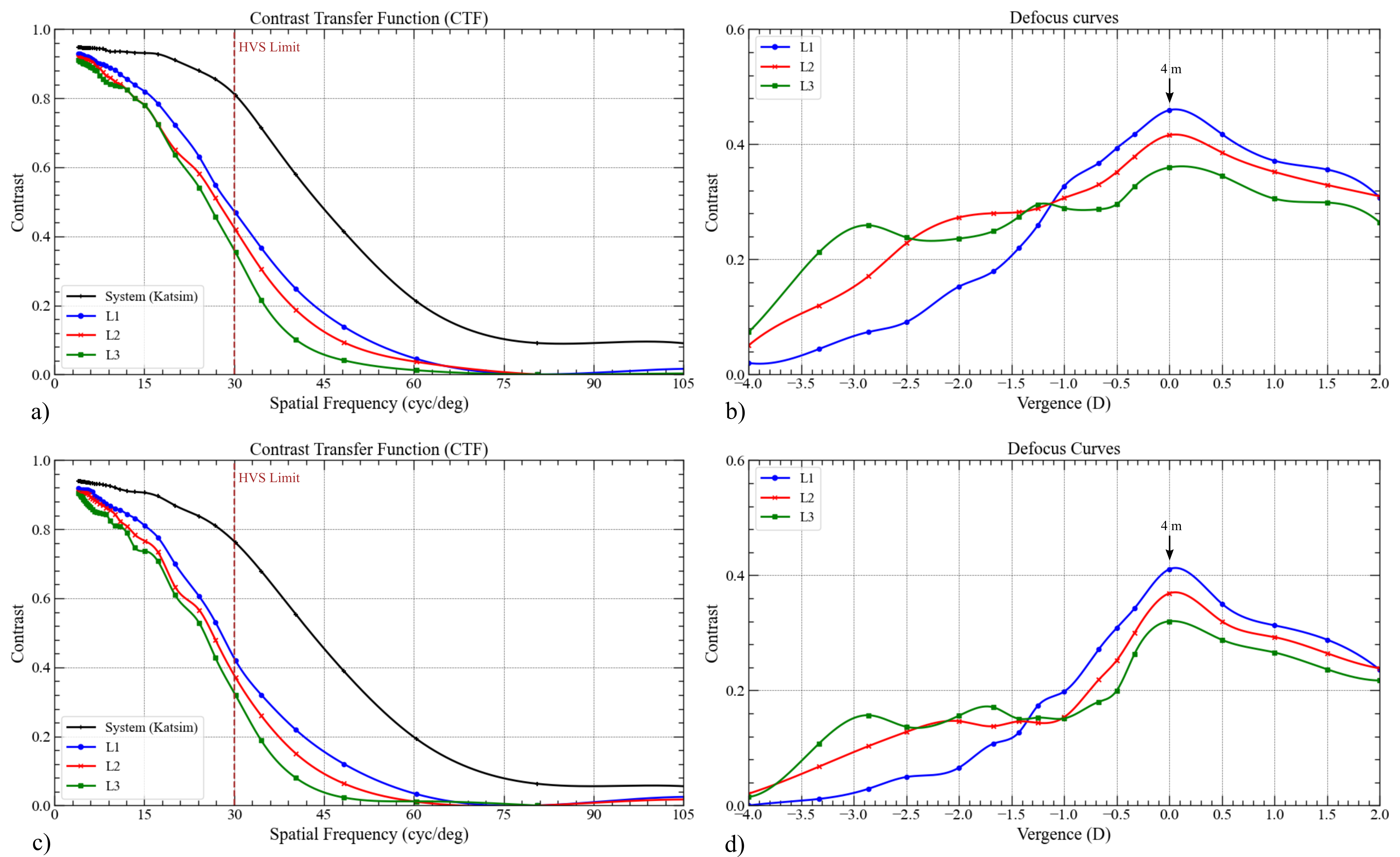}
\caption{Experimental results for CTF and defocus‑curve analysis.
(a) and (c) CTF results for the small (400\,\textmu m) and large (2.2 mm) illumination‑source sizes, respectively, using customized spatial‑frequency content. The red dashed line marks the human‑visual‑system (HVS) resolution limit at 30 (cyc/deg).
(b) and (d) Defocus‑curve analyses at 30 (cyc/deg) for the same small and large illumination‑source sizes, respectively; here 0 D corresponds to a virtual‑image distance of 4 m. Intraocular‑lens types: L1 = Monofocal, L2 = EDOF, L3 = Trifocal.} 
\label{fig:8}
\end{figure}

\section{Discussion}
The present results demonstrate that synchronizing an AM-SLM with a high-speed varifocal lens and RGB time-multiplexed illumination yields stable multi-plane retinal imagery while decoupling focus control from eyebox geometry. Within this architecture, the 60 fps SLM with RGB subframes provides an effective 180 Hz depth-coded cycle across a tunable vergence range (0.2 m to optical infinity), and the illumination-defined eyebox (1-5 mm) preserves the spatial origin and angular magnification as focus is swept. These properties enable Katsim to support both perceptual simulation for patients and quantitative optical measurements for lenses without reconfiguration.

The contrast-transfer analyses in Figure 8, obtained by displaying calibrated spatial-frequency targets (Figure 7) and capturing the resulting images through an artificial eye, indicate that the display pipeline itself sustains high contrast well into mid-to-high spatial frequencies, well beyond the typical HVS limit marked at ~30 cyc/deg, and therefore attenuation in lens-specific curves can be attributed to the IOLs rather than display-induced artifacts. This is consistent with the system design, where depth coding and eyebox shaping are controlled independently and the optical relay is kept fixed while diopter is modulated electronically.

Interpreting the lens-dependent outcomes, L1 (monofocal) achieves the highest peak contrast at best focus but exhibits a steep fall-off with defocus, reflecting a single-focus design. L2 and L3 (EDOF/trifocal) broaden the through-focus response, maintaining useful contrast over a wider vergence span and, at times, forming multiple shoulders, at the expected cost of peak contrast relative to L1. Across CTFs, the curves converge at low spatial frequencies and diverge at higher frequencies ( $\approx 40\text{--}60~\mathrm{cyc/deg}$ ), where L1 more often retains contrast while L2/L3 roll off more rapidly.

Reducing the illumination size ($\approx 400~\mu\mathrm{m}$; Fig. 8a,b) acts like a smaller effective pupil, trimming off-axis rays and thereby lowering aberrations. As expected, the CTF shows a gentler high-frequency roll-off and higher peak contrast. Conversely, enlarging the light source size ($\approx 2.2~\mathrm{mm}$; Fig. 8c,d) increases angular diversity, which enhances the separation of focal zones in the defocus curves (more distinct peaks for multifocal optics) but lowers peak contrast across vergences due to aberration-related blur. In the trifocal lens, this manifests as relatively similar secondary-peak amplitudes with the small size (less focal stratification; the three foci appear more even) versus more uneven peak heights with the large size (greater focal stratification, i.e., clearer depth separation, but reduced absolute contrast). EDOF behavior follows the same principle: a broad through-focus plateau is more contrast-preserving for the small size and becomes more clearly differentiated, but overall lower in amplitude, for the large size. Because pupil diameter varies with luminance, these two operating regimes provide a practical proxy for photopic (smaller pupil/daylight) and mesopic (larger pupil/dim-light) conditions relevant to patient counseling and standardized optical-bench characterization.

These findings have practical implications for preoperative counseling and IOL selection. For patients prioritizing peak acuity at a single distance, monofocal behavior (as seen in L1) remains advantageous; for those valuing functional range (e.g., near-to-intermediate tasks), the broadened defocus characteristics of EDOF or trifocal optics (L2/L3) may better align with needs despite reduced peak contrast. Because Katsim’s pupil-tracked steering can route imagery through clear, non-cataractous regions of the crystalline lens in real time, the platform can emulate patient-specific aperture conditions (e.g., irregular pupils or sectoral opacity) when previewing these trade-offs under photopic and mesopic scenarios.

Methodological considerations include the present FOV ($\approx 9.15^{\circ}$), the use of three time-multiplexed depth planes per frame cycle, and sequential RGB illumination. While 180 Hz depth coding limits perceptual cross-talk, increasing the number of planes or FOV may require further optimization of SLM timing, lens settling dynamics, and illumination duty cycle. The artificial-eye/bench configuration, essential for repeatable CTF/defocus measurements, does not capture accommodation dynamics, tear-film variability, or neural adaptation present in vivo; psychophysical validation on healthy and cataract cohorts will therefore be necessary to map bench metrics to perceived image quality. Finally, absolute contrast depends on spectral bandwidth, alignment tolerances, and camera/retina sampling; these were controlled during experiments but represent standard sources of uncertainty for any imaging-based MTF/CTF assessment.

\section{Conclusion}
The proposed device is a compact, modular near-eye display that unifies patient-specific visual simulation and objective IOL evaluation. By synchronizing an AM-SLM, RGB illumination, and a high-speed varifocal lens, the system renders full-color, multi-depth scenes at an effective 180 Hz while maintaining a fixed image origin and a configurable 1–5 mm eyebox across a 0.2 m–optical infinity depth range (FOV $\approx 9.15^{\circ}$). Bench experiments confirm that the display pipeline preserves high contrast, allowing measured differences in CTF/defocus behavior to be attributed to the IOLs rather than system artifacts. These capabilities position the platform as both a counseling tool for personalized preoperative planning and a research-grade system for rigorous IOL characterization. With forthcoming expansions in FOV, plane count, and in-vivo validation, the proposed device can help align clinical decision-making with quantitative optics and accelerate innovation in IOL design and evaluation.

\begin{backmatter}

\bmsection{Funding}
This study has been supported by the European Innovation Council (EIC) Transition program, funded by the European Union (EU) (support agreement number 101057672) and TUBITAK 2247 program (Turkey) (support agreement number 120C145).

\bmsection{Acknowledgment}
We would like to thank Dr. Goksen Yaralioglu, Evan Carter, and Dr. Yasser Elkahlout from CY Vision, as well as Selim Olcer and Fırat Turkkal from the Optical Microsystems Laboratory (OML), Koç University, for their valuable assistance with the electronic and mechanical design of the devices used in these setups.

\bmsection{Disclosures}
The authors declare no conflicts of interest.

\bmsection{Data availability} Data underlying the results presented in this paper are not publicly available at this time but may be obtained from the authors upon reasonable request.






\end{backmatter}

\bibliography{references}

\begin{thebibliography}{10}
\newcommand{\enquote}[1]{``#1''}

\bibitem{bourne2013causes}
R.~R. Bourne, G.~A. Stevens, R.~A. White, \emph{et~al.}, \enquote{Causes of vision loss worldwide, 1990--2010: a systematic analysis,} {\protect\JournalTitle{The lancet global health}} \textbf{1}, e339--e349 (2013).

\bibitem{marcos2021simulating}
S.~Marcos, E.~Martinez-Enriquez, M.~Vinas, \emph{et~al.}, \enquote{Simulating outcomes of cataract surgery: important advances in ophthalmology,} {\protect\JournalTitle{Annual review of biomedical engineering}} \textbf{23}, 277--306 (2021).

\bibitem{soomro2024head}
S.~R. Soomro, S.~Sager, A.~M. Paniagua-Diaz, \emph{et~al.}, \enquote{Head-mounted adaptive optics visual simulator,} {\protect\JournalTitle{Biomedical Optics Express}} \textbf{15}, 608--623 (2024).

\bibitem{marx2024ability}
S.~Marx, O.~Kolbe, M.~Gerlach, \emph{et~al.}, \enquote{The ability of a virtual implantation device to evaluate two intraocular lens designs,} {\protect\JournalTitle{Journal of Refractive Surgery}} \textbf{40}, e911--e915 (2024).

\bibitem{aydindougan2020applications}
G.~Ayd{\i}ndo{\u{g}}an, K.~Kavakl{\i}, A.~{\c{S}}ahin, \emph{et~al.}, \enquote{Applications of augmented reality in ophthalmology,} {\protect\JournalTitle{Biomedical optics express}} \textbf{12}, 511--538 (2020).

\bibitem{barcala2023visual}
X.~Barcala, A.~Zaytouny, D.~Rego-Lorca, \emph{et~al.}, \enquote{Visual simulations of presbyopic corrections through cataract opacification,} {\protect\JournalTitle{Journal of Cataract \& Refractive Surgery}} \textbf{49}, 34--43 (2023).

\bibitem{marcos2022adaptive}
S.~Marcos, P.~Artal, D.~A. Atchison, \emph{et~al.}, \enquote{Adaptive optics visual simulators: a review of recent optical designs and applications,} {\protect\JournalTitle{Biomedical optics express}} \textbf{13}, 6508--6532 (2022).

\bibitem{akyazi2024intraocular}
D.~Akyazi, U.~Aygun, A.~Sahin, and H.~Urey, \enquote{Intraocular lens simulator using computational holographic display for cataract patients,} {\protect\JournalTitle{Plos one}} \textbf{19}, e0295215 (2024).

\bibitem{kavakli2021pupil}
K.~Kavakl{\i}, G.~Ayd{\i}ndo{\u{g}}an, E.~Ulusoy, \emph{et~al.}, \enquote{Pupil steering holographic display for pre-operative vision screening of cataracts,} {\protect\JournalTitle{Biomedical Optics Express}} \textbf{12}, 7752--7764 (2021).

\bibitem{na2022novel}
K.-S. Na, S.-J. Kim, G.~Nam, \emph{et~al.}, \enquote{A novel intraocular lens simulator that allows patients to experience the world through multifocal intraocular lenses before surgeries,} {\protect\JournalTitle{Translational Vision Science \& Technology}} \textbf{11}, 14--14 (2022).

\bibitem{benedi2021optical}
C.~Benedi-Garcia, M.~Vinas, C.~M. Lago, \emph{et~al.}, \enquote{Optical and visual quality of real intraocular lenses physically projected on the patient’s eye,} {\protect\JournalTitle{Biomedical Optics Express}} \textbf{12}, 6360--6374 (2021).

\bibitem{lanman2013near}
D.~Lanman and D.~Luebke, \enquote{Near-eye light field displays,} {\protect\JournalTitle{ACM transactions on graphics (TOG)}} \textbf{32}, 1--10 (2013).

\bibitem{liu2023recent}
S.~Liu, Y.~Li, and Y.~Su, \enquote{Recent progress in true 3d display technologies based on liquid crystal devices,} {\protect\JournalTitle{Crystals}} \textbf{13}, 1639 (2023).

\bibitem{zaytouny2023clinical}
A.~Zaytouny, I.~Siso-Fuertes, X.~Barcala, \emph{et~al.}, \enquote{Clinical validation of simulated multifocal intraocular lenses. simvis gekkotm simulations and reports on pseudophakic patients,} {\protect\JournalTitle{Investigative Ophthalmology \& Visual Science}} \textbf{64}, 5421--5421 (2023).

\bibitem{paniagua2023economic}
A.~M. Paniagua-Diaz, J.~Mompean, S.~Shomroo, and P.~Artal, \enquote{Economic and compact modulation unit for visual simulation: the advantages of vertical aligned liquid crystal devices,} in \emph{Ophthalmic Technologies XXXIII,}  vol. 12360 (SPIE, 2023), pp. 91--92.

\bibitem{krosl2020cataract}
K.~Kr{\"o}sl, C.~Elvezio, L.~R. Luidolt, \emph{et~al.}, \enquote{Cataract: Simulating cataracts in augmented reality,} in \emph{2020 IEEE International Symposium on Mixed and Augmented Reality (ISMAR),}  (IEEE, 2020), pp. 682--693.

\bibitem{jones2020seeing}
P.~R. Jones, T.~Somoske{\"o}y, H.~Chow-Wing-Bom, and D.~P. Crabb, \enquote{Seeing other perspectives: evaluating the use of virtual and augmented reality to simulate visual impairments (openvissim),} {\protect\JournalTitle{NPJ digital medicine}} \textbf{3}, 32 (2020).

\bibitem{wu2025using}
Z.~Wu, A.~Zieli{\'n}ska, G.~U. Auffarth, \emph{et~al.}, \enquote{Using a look-through simulator to analyze the correlation between functional and optical properties of intraocular lenses,} {\protect\JournalTitle{Journal of the Optical Society of America A}} \textbf{42}, C6--C12 (2025).

\bibitem{zolfaghari2025bright}
P.~Zolfaghari, F.~O. Ozhan, and H.~Urey, \enquote{Bright pupil-based pupil center tracking using a quadrant photodetector,} {\protect\JournalTitle{Optics \& Laser Technology}} \textbf{186}, 112762 (2025).

\bibitem{akyazi2023artificial}
D.~Akyazi, K.~Kavakli, U.~Aygun, \emph{et~al.}, \enquote{Artificial eye model and holographic display-based iol simulator,} in \emph{Ophthalmic Technologies XXXIII,}  vol. 12360 (SPIE, 2023), pp. 177--182.

\end{thebibliography}
\end{document}